\documentclass[sn-mathphys,Numbered]{sn-jnl}
\usepackage{graphicx}%
\usepackage{multirow}%
\usepackage{amsmath,amssymb,amsfonts}%
\usepackage{amsthm}%
\usepackage{mathrsfs}%
\usepackage[title]{appendix}%
\usepackage{xcolor}%
\usepackage{textcomp}%
\usepackage{manyfoot}%
\usepackage{booktabs}%
\usepackage{algorithm}%
\usepackage{algorithmicx}%
\usepackage{algpseudocode}%
\usepackage{listings}%

\begin{document}

\title[Article Title]{A novel floating piezoelectric energy harvesting from water waves: fully-coupled simulation}

\author[1]{Erfan Rajabi Khorramdarreh}
\author[2]{Mohammad Mostafa Mohammadi}\email{dr.mohammadi@znu.ac.ir}
\author[1]{Jafar Ghazanfarian}
\affil*[2]{Mechanical Engineering Department, Faculty of Engineering, University of Zanjan, Zanjan, Iran}
\affil[1]{Mechanical Engineering Department, Faculty of Engineering, University of Zanjan, Zanjan, Iran}

\abstract{A fully-coupled-fluid-structure-piezoelectric model is presented based on the finite element method that is capable of modeling piezoelectric harvesters in the presence of free-surface flow and floating lightweight harvesters with arbitrary movements. The Navier-Stokes equations and the phase-field method are employed to describe the free-surface waves. Equations of the conservation of linear momentum in company with the piezoelectric constitutive relations in the strain-charge form are utilized to obtain solid deformation and the electric field intensity. According to the results, attaching mass to the tip of the beam leads to 13.5\% rise in the output voltage compared to the state without the attached mass. Another studied factor was the influence of the load resistance on voltage and the output power. The generated voltage grows along with the load resistance until it reaches a constant value. However, the power has an optimum load resistance that is 2.61 times higher than the reference state. The beam's inclination is significant in effectively exploiting water waves due to raising the root mean square (RMS) value of the voltage  by 89.53\% at an angle of 40 degrees relative to the vertical state. By altering the thickness of the beam from 1 mm to the value of 1.5 mm, the RMS voltage exhibits a considerable upward change of 66\%.  By increasing the length of the cantilever beam connected to the buoyant structure, and therefore, the indentation of the beam in the water, the output voltage grows, such that a beam with the length of 25 cm shows a 2.92 times increase in the output voltage relative to the beam with a length of 15 cm.}


\maketitle




\section{Introduction}
\label{Intro}
Due to ongoing advancements in microcontrollers and microelectronics, designing sensors with lower energy consumption is receiving much attention nowadays, and developments in these devices have outpaced conventional battery technologies \cite{Kim, aabid2021systematic}. Using new energy harvesting techniques to acquire electricity from the environment is a viable answer to the existing problems. Water flow is ubiquitous in nature and in industries from small to large scales. Also, they have higher power density potential in comparison with other sources of renewable energy like solar and wind energy.
Water waves give an opportunity to harness energy from the kinetic energy of flow through mechanical vibrations due to wave-instruction interaction \cite{kim2021wave, lass2019rotor}.

Three mechanisms of converting mechanical vibrations to electrical energy have been proposed: electrostatic \cite{mitcheson2004mems}, electromagnetic transducers \cite{glynne2004electromagnetic}, and piezoelectrics \cite{toprak2014piezoelectric}. The prime features of piezoelectric materials are their great power density (generated power with respect to volume), lower cost of fabrication and simplicity of use along with their versatility in a wide range of structures and applications \cite{kiran2020progress}.

Accompanying Piezoelectric technology with the fluid-solid interaction (FSI) phenomenon is a viable alternative for small-scale turbines because of its lower cost and simpler mechanism \cite{elhami2021vibration}. When water waves hit an elastic structure, the imposed mechanical vibration could be converted into electricity via attached piezoelectric materials to the structure. One of the most common piezoelectric materials is PZT (lead zirconate titanate) materials, which contributes to a large energy conversion rate owing to their high electro-mechanical coupling constants \cite{erturk2008distributed}.

In recent years, many researchers have studied piezoelectric harvesters buried in fluid flow. Thus, lots of models and experiments have been designed to analyze the system’s behavior. The following lines will be dedicated to reviewing numerical approaches adopted by scholars for studying piezoelectric energy harvesting from water flows.

Hu et al. \cite{hu2018modeling} proposed an analytical model as well as an experimental examination of energy harvesting via vortex-shedding vibrations. The vortices shedding on the vibrating plate and their flow behaviour has been considered by the Lamb-Oseen vortex model to provide a simpler physical grasp of the problem. Then, they explored the impact of the Reynolds number and finding optimum distance for voltage and power generation. Nabavi et al. \cite{nabavi2019ocean} introduced an innovative piezoelectric-based ocean waves energy convertor that can generate power from breaking ocean waves on a vertical face. They also used classical beam theory accompanied by Lagrange’s method to derive electro-mechanical formulations analytically and then validated the results by experimental data.

Sang et al. \cite{song2015study} researched harvesting energy from vortices induced by water flow on a cantilever harvester with a cylindrical blade attached to its tip. They proved that expanding the diameter of the cylinder and decreasing its mass resulted in higher voltage by performing an experiment and providing a mathematical model using the Euler-Bernoulli equation, which is a distributive parametric model. An efficient piezoelectric coupled buoy was created by Wu et al. \cite{wu2015ocean} to harvest energy from ocean waves. Several cantilever beams containing a piezoelectric material were joined to a buoyant structure, allowing it to be readily suspended in intermediate and deep ocean waves. They provided a numerical model using Euler-Bernoulli's beam theory, which describes and accounts for forces acting on the buoyant object including the drag, the lift, and the buoyancy forces.

Guo et al. \cite{gu2020effects} conducted a two-dimensional numerical simulation to explore performance of the VIV harvester's installation depth while considering it as a mass-spring-damper oscillator and solving governing equations by applying the Newmark-$\beta$ approach through a UDF code. The simulation findings revealed that if the installation depth is more than half a meter, the free-surface of the flow has no effect on the harvester's performance.  Xie et al. \cite{xie2014potential} developed a piezo-beam harvester comprised of a cantilever beam with an attached mass to generate power from the longitudinal waves. They provided a mathematical model for the resistance and generated voltage by means of the linear theory of Airy wave and classical elastic beams in order to investigate the influence of factors like the attached mass to the beam tip, the ratio of the beam width to the beam thickness, and the wave height. By adding weight to the tip and raising the mass proportion of weight to the mass of the beam, the output voltage could be up to 6 times higher than the case without the attached weight.

In order to examine the generated voltage of various piezoelectric energy harvester (PEH) arrangements found on vortex-induced oscillations, that is typically utilized in low-velocity water in most of surroundings, Sui et al. \cite{sui2022study} performed experiments and simulations by considering harvesters as a single-degree-of-freedom (SDOF) model, while Ramirez \cite{ramirez2021coupled} demonstrated the effect of three cylinders operating in tandem arrangement on the voltage. Yayla et al. \cite{yayla2020case} used a similar numerical strategy, embedding a piezoelectric generator behind
vortex-generating plates involving nozzles of various angles and diameters in a single-phase turbulent flow channel. Yamac et al. \cite{yamacc2021numerical} simulated a wave tank by employing the moving mesh concept and the volume-of-fluid technique along with a code containing a similar methodology in Wu et al.'s article to assess the potency of fluid column height and waves height.

Using magnets and flow-induced vibrations, Zou et al. \cite{zou2021magnetically} developed bistable piezoelectric harvesting process. Piezoharvester magnets are influenced by driving magnets on cantilever beams. The system is defined as a mass-spring-damper oscillator. The blade angle and water velocity affect the output power. As the water velocity rises, so does the driving magnet displacement.
Higher hydrodynamic area and vibration and power are results of an increased blade angle. Viet et al. \cite{viet2016energy} developed a mass-spring system with piezoelectricity-lever devices for vibration amplification to transform wave motion to mechanical vibration through a floater. A mathematical model stemmed from the Lagrangian-Euler approach is proposed and resolved using the iterative technique to determine the root-mean-square value of the output electric power in order to study factors such as the moving mass in the system and the ocean wave amplitude. Mirab et al. \cite{mirab2015energy} employed numerical optimization for the variables that play role in harnessing electricity assisted with taking into account methods of the linearized wave theory and JONSWAP spectra formula to enhance the effective electric power generated from a PZT cantilever harvester restricted at the ocean floor.

Most of the aforementioned numerical works have modeled the energy harvesting system for single-phase flows, and in order to accomplish this, the piezoelectric beam is considered as a single-degree-of-freedom system. However, this method is ineffective at higher working frequencies. Furthermore, in simulations related to floating harvesters surrounded by the free-surface flow of water, several constraints are imposed on their movement so that  the methods like finite-element or distributed parameter model can be implemented. These constraints do not apply to lightweight floaters. As a consequence, a fully-coupled simulation of the piezoelectric beam with the two-phase flow (free-surface flow of air-water) could address the mentioned problems. To the best of author's knowledge, this is the first attempt to investigate a fully-coupled simulation of a floating piezoelectric beam in a free-surface wavy flows. The current study intends to provide a fluid-structure-piezoelectric model established upon the finite element method in COMSOL software in order to simulate piezoelectric harvesters interacting with water waves. The physical conceptualization was established by coupling the Navier-Stokes equations for incompressible fluid flow, the Gauss law for circuit and piezoelectricity equations, and the Navier equations for characterizing the deformation of the elastic beam. The final results is to develop the design of a harvester floating over water waves as our main engineering problem, which is so important for remote area and offshore applications.

The following structure best describes the article. After the literature review, the configuration feature and the governing equations of a piezoelectric harvesting system established on water waves are represented in  section \ref{method}. Sections \ref{very} and \ref{results}  contains the numerical verification and results, respectively. Finally, section \ref{deduction} demonstrates the conclusions.

\section{Governing equations}
\label{method}
\subsection{Problem statement and experimental setup}
\begin{figure}[htbp]
     \centering
     \includegraphics[width=12.5 cm]{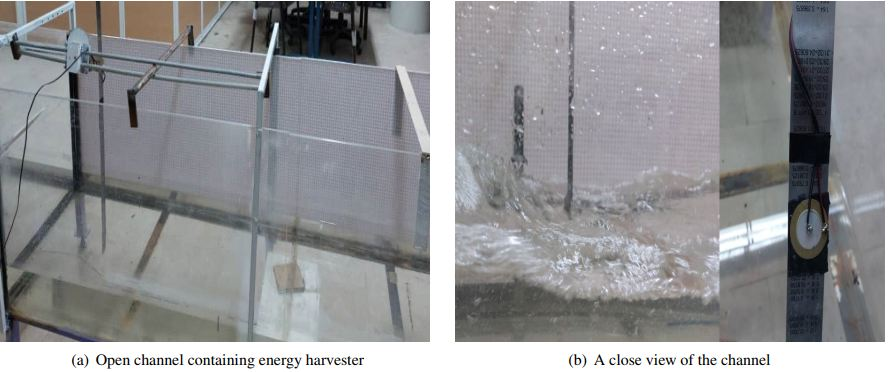}
     \caption{Configuration of the energy harvesting system.}
     \label{fig:Geometry}
\end{figure}

\begin{figure}[htbp]
     \centering
     \includegraphics[width=11.5 cm]{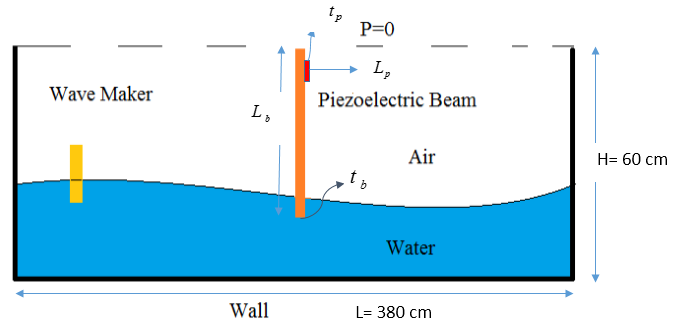}
     \caption{Front view of the 2D geometric model used in the simulations.}
     \label{fig:schematic of geometry}
\end{figure}

The harvesting system under study is shown in Fig. \ref{fig:Geometry}. The system is comprised of a stainless steel beam and a piezoelectric layer of PZT-4 mounted on one side of the beam (unimorph piezoelectric energy harvester). Surface water waves created by a piston-type wavemaker provide time-varying force on the cantilever beam, which is needed for the piezoelectric layer to generate voltage. This layer is paralleled with an electrical resistance and located near the beam's support due to being exposed to a higher range of imposed stresses in comparison to other distances. Fig. \ref{fig:schematic of geometry} and Tab. \ref{tab: geo info} present the schematic geometry of the problem and geometrical parameters, respectively.

\begin{table}[h]
\caption{\label{tab: geo info} Geometrical properties of the energy harvesting system.}
\begin{tabular}{ c p{4cm} c  }
 \hline
 Parameter & Description &  Value \\
 \hline
 H & Height of the channel &  60 (cm) \\
 L & Length of the channel &   380 (cm) \\
  ${L_b}$ & Length of the beam &  52 (cm) \\
 ${t_b}$ & Thickness of the beam & 1 (mm)  \\
 ${L_p}$ & Length of the piezoelectric & 2.54 (cm)\\
 ${t_p}$ & Thickness of the piezoelectric & 0.1 (mm)\\
 R & Electrical resistance &  1 (M$\Omega$) \\
 ${I_W}$ & Wavemaker indentation in water &  4 (cm) \\
  ${I_b}$ & Beam indentation in water &  3 (cm) \\
  $\omega $ & Frequency of wave maker &  1.37 (Hz) \\
 \hline
\end{tabular}
\end{table}

The wavemaker, by reciprocating motion, creates a turbulent flow in the channel. The excitation frequency of the surface waves is small. Also, concerning the material type of the beam, the use of linear elastic model is reasonable to simulate the harvesting process. Furthermore, it is supposed that the beam has no twist and torsional shear stress; thus, it is reasonable to simulate the problem in two dimensions. Due to the multiphysics essence of the system, it is vital to reduce the computational cost.

Water and air are considered as Newtonian fluids that are governed by continuity and momentum equations in unsteady and ALE (arbitrary Lagrangian–Eulerian) form\cite{murea2010arbitrary}:
\begin{equation}
\frac{{\partial \rho }}{{\partial t}}\; + \;\nabla .\,(\rho u) = 0
\end{equation}
 \begin{equation}
\rho \frac{{\partial \,u}}{{\partial \,t}} + \rho \,((u-w){\kern 1pt} .\nabla u) = \nabla .{\kern 1pt} \,[ - pI + K] + F
\end{equation}
where u and p are the fluid velocity and pressure, w is the mesh velocity, $\rho$ is the density and F is the external force. The standard $k - \varepsilon$ model is applied to capture turbulence generated in the channel, which is commonly used to study wave simulations in such applications \cite{karim2014numerical}.
\begin{equation}
K = (\mu  + {\mu _t})(\nabla u + \nabla {u^T})
\end{equation}
\begin{equation}
\rho \frac{{\partial \,\kappa }}{{\partial \,t}} + \rho (u\,.\nabla )\kappa  = \nabla .[(\mu  + \frac{{{\mu _t}}}{{{\sigma _k}}})\nabla \kappa ] + {P_k} - \rho \varepsilon
\end{equation}
\begin{equation}
\rho \frac{{\partial \,\varepsilon }}{{\partial \,t}} + \rho (u\,.\nabla )\varepsilon  = \nabla .[(\mu  + \frac{{{\mu _t}}}{{{\sigma _\varepsilon }}})\nabla \kappa ] + {C_{\varepsilon 1}}\frac{\varepsilon }{\kappa }{P_\kappa } - {C_{\varepsilon 2}}{\kern 1pt} \rho \frac{{{\varepsilon ^2}}}{\kappa }
 \end{equation}
\begin{equation}\label{eq:6}
 {\mu _t} = \rho \,{C_\mu }\frac{{{\kappa ^2}}}{\varepsilon }
\end{equation}
\begin{equation}
{P_\kappa } = {\mu _t}[\nabla u:(\nabla u + (\nabla {u^T})]
 \end{equation}
The turbulent viscosity $\mu _t$  is calculated by k and $\varepsilon$ in Eq. (\ref{eq:6}), where $C_\mu$ is an empirical coefficient whose value is often deemed to be 0.09. The model constants are $ {C_{1\varepsilon }} = 1.44\quad \;{C_{2\varepsilon }} = 1.92\quad \;{\sigma _k} = 1\quad \;{\sigma _\varepsilon } = 1.3\quad \;$ \cite{franco2023numerical}.

The computational grid must be deformed throughout time to capture the changes in the flow caused by body deflection. In spite of the ever-growing popularity of the space-time methods in FSI problems, this work resorted to the ALE methods which are straightforward to implement. The Laplace's equation governs the mesh movement.
\begin{equation}
    {\nabla ^2}w = 0\;\;,\;\,on\;{\Omega _f}\quad \;\;\;\;w = \frac{{\partial \eta }}{{\partial t}}\;,\quad on\sum \,(t)
\end{equation}
In the equation above, $w$ is the grid velocity, and $\eta$ denotes the displacement of the fluid-solid interface ($\Sigma \,(t)$). To update the grid's point for reference and the current mesh deformation, Eq. (\ref{configuration }) is used \cite{wang2020numerical, huang2010adaptive}:
\begin{equation}\label{configuration }
  {x^{n + 1}} - {x^n} = r\Delta t
\end{equation}

For smoothing the dynamic mesh, the Winslow method yields one coupled system of equations for all coordinate directions, which is non-linear and inclined to permit elements to be stretched excessively. Thus, large displacements before inverting are allowable \cite{knupp1999winslow}.

\subsection{The phase-field method}
\label{phase}
In order to track the air/water interface (two immiscible fluids), a diffuse-interface technique named the phase-field method is applied to capture the exact position of the interface by mixing the free energy. This method can handle complicated morphological changes in the interface, and does not necessitate calculating interfacial curvature, which is usually challenging. An auxiliary variable $\phi (x,t)$ is employed to trace the air-water interface. $\phi $ = 1 and $\phi $ = -1 present the pure air and pure water, respectively. This function changes smoothly around the interface between 0 and 1. The dependence of viscosity and density on the phase-field function leads the Navier-Stokes equations coupled with the convective Cahn-Hilliard equation describing the evolution of the phase-field function \cite{yue2006phase, teixeira2022heat, akter2017computational}. The Cahn-Hilliard equation in the ALE framework is as follows:
\begin{equation}\label{eq:cahn}
\frac{{\partial \,\phi }}{{\partial \,t}} + (u-w)\,.\nabla \phi  = \nabla .\,\gamma \,\nabla G
\end{equation}
\begin{equation}
G = \lambda \,[ - {\nabla ^2}\phi  + \frac{{\phi \,({\phi ^2} - 1)}}{s}]
\end{equation}
where G and $\phi $  are chemical potential and the dimensionless phase-field function, respectively. The phase-field interface decomposes the fourth-order PDE in Eq. (\ref{eq:cahn}) into two second-order PDEs:
\begin{equation}
\frac{{\partial \,\varphi }}{{\partial \,t}} + (u-w).\nabla \phi  = \nabla .\frac{{\lambda \gamma }}{{{s ^2}}}\nabla \psi
\end{equation}
\begin{equation}
\psi  =  - \nabla .\,{s ^2}\nabla \phi  + {\phi ^{}}\,({\phi ^2} - 1)
\end{equation}
where $\gamma$, $\lambda$, $s$ symbolize the mobility of the interface, the mixing energy density, and a controlling interface parameter, respectively. Half of the largest mesh size passing through the interface was prescribed as a controlling parameter. The mobilizing parameter sets the time scaling for the Cahn-Hilliard equation. It should be both enough large or small to keep a fixed thickness in the interface while preventing overly damping of the convective terms.  $\gamma$ has an empirical value for the mobility of the interface in the Cahn-Hilliard equation. Dong and Shen \cite{dong2012time} presented details about the influence of $\gamma$ in their tests, finding that when its value is less than $10{\,^{ - 4}}$, it has no impact on the outcomes of the capillary wave study. In this work, given the maximum surface velocities created by the wavemaker, the value of $5 \times {10^{\, - 5}}$ is considered \cite{zheng2021fluid}. The following Eqs. (\ref{coupling}) connect the dynamic viscosity, $\mu$, and the density, $\rho$, to the phase-field variable.
\begin{equation}\label{coupling}
    \rho (\phi ) = \frac{{{\rho _1} + {\rho _2}}}{2} + \frac{{{\rho _1} - {\rho _2}}}{2}\phi \;\quad ,\quad \mu (\phi ) = \frac{{{\mu _1} + {\mu _2}}}{2} + \frac{{{\mu _1} - {\mu _2}}}{2}\phi
\end{equation}

Imposing initial and boundary conditions on the fluid domain is of great significance. One of these boundary conditions is the continuity of velocities on the solid interface; the other represents a wetted wall. When the waves intersect the walls in the computational domain, it is required to specify the contacting angle, ${\theta _w}$. For simplicity, a contact angle of ${90^o}$ was appointed \cite{huang2015wetting}. Also, the mass flow across the walls equals zero:
\begin{equation}
n.\,\frac{{\lambda \gamma }}{{{s ^2}}}\nabla \psi  = 0
\end{equation}
To specify the initial position of the boundary between two fluids, the phase-field variable could be specified explicitly for a time-dependent study to be run directly. The initial conditions are given by:
\begin{equation}
u(x,0) = {u_0}\,(x)\quad ,\quad \phi (x,0) = {\phi _0}\,(x)
\end{equation}

\subsection{Structure}
\label{solid}
The  linear momentum conservation equations govern the structure deformation. The solid problem describes the beam dynamics through an isotropic linear elastic model:
\begin{equation}
{\rho _s}\frac{{{\partial ^2}u_s}}{{\partial \,{t^2}}} = \nabla .\,{\sigma _s} + {\rho _s}{b_s}
\end{equation}
where $u_s$ is the solid displacement vector, $\rho _s$ denotes the solid density and $\sigma _s$ is the Cauchy stress tensor, which is defined as follows:
\begin{equation}
{\sigma _s} = c:\varepsilon_s
\end{equation}
\begin{equation}\label{strain tensor}
\varepsilon_s  = \frac{1}{2}[\nabla u _s + (\nabla u_s){{\kern 1pt} ^T}]
\end{equation}
in which c is the linearized Young's modulus, and $\varepsilon_s$ is the strain tensor obtained according to Eq. (\ref{strain tensor}).

The fluid motion equations, the structure deformations laws, and the transmission conditions at the fluid-structure interface characterize a fluid-solid interaction. The solid-fluid interface was assigned kinematic and dynamic constraints to form the FSI coupling \cite{molinos2023derivation}.
\begin{equation}\label{kinematic}
\frac{{\partial \,\eta }}{{\partial \,t}} = u
\end{equation}
\begin{equation}\label{dynamic}
{\sigma _f}.\,{\kern 1pt} n + {\sigma _s}.{\kern 1pt} \,n = 0
\end{equation}
Eq. (\ref{kinematic}) shows the equity of solid and water velocity at the interface (also implies the no-slip boundary condition). In addition, Eq. (\ref{dynamic}) describes the force balance at the interface. $\eta$ and $n$ are the interface displacement and the outward normal vector of the solid boundary, respectively.

\subsection{The piezoelectric effect}
\label{piezo}
Piezoelectricity is a property of specific crystalline materials that could convert an applied load to electricity. In other words, it is a coupled behaviour of strain change and electrical charge due to an applied electric and stress fields. The piezoelectric poled through the thickness could be modeled by two linearized constitutive equations in the strain-charge form:
\begin{equation}
\varepsilon_s  = {s_e}\,{\sigma _s} + {d^T}{E_p}
\end{equation}

\begin{equation}
D = d\,{\sigma _s} + {\varepsilon _T}\,{E_p}
\end{equation}
where $D$ is the electric displacement field, $E$ is the electric field, and $\varepsilon _T $ is the dielectric permittivity. $d$, $d^T$, $s_e$ are the elastic compliance matrix, the piezoelectric coefficient, and the piezoelectric constants, respectively. In order to complete the electro-mechanical coupling, it is necessary to relate the piezoelectric equations with the circuit through an electrostatic module. An electrostatic module is not a time-dependent physical aspect unless coupled with other modules. Through Gauss's law, this coupling is possible. The piezoelectric thickness is subjected to the zero charge boundary condition, while the piezoelectric-beam interface has the ground boundary condition. Furthermore, the other side of the material is bounded by the terminal boundary condition to  obtain the generated voltage by the PZT material through a paralleled load resistance \cite{mishra2020analytical, fan2022energy}.

\subsection{Grid/time-step size independence tests}
\label{mesh}
\begin{figure}[h]
\centering
\includegraphics[width=8.5cm]{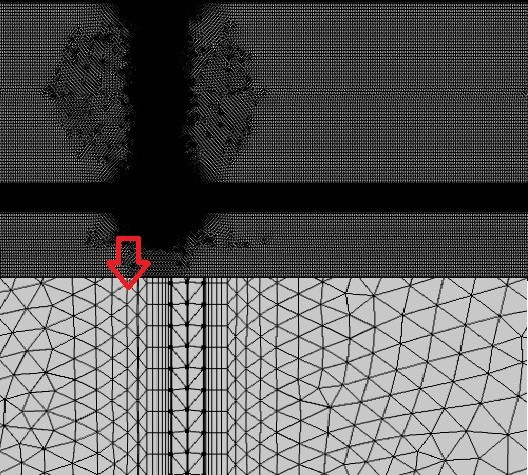}
\caption{\label{fig:mesh}The computational grid and a close snapshot of the beam.}
\end{figure}

\begin{figure}[t]
\centering
\includegraphics[width=10cm]{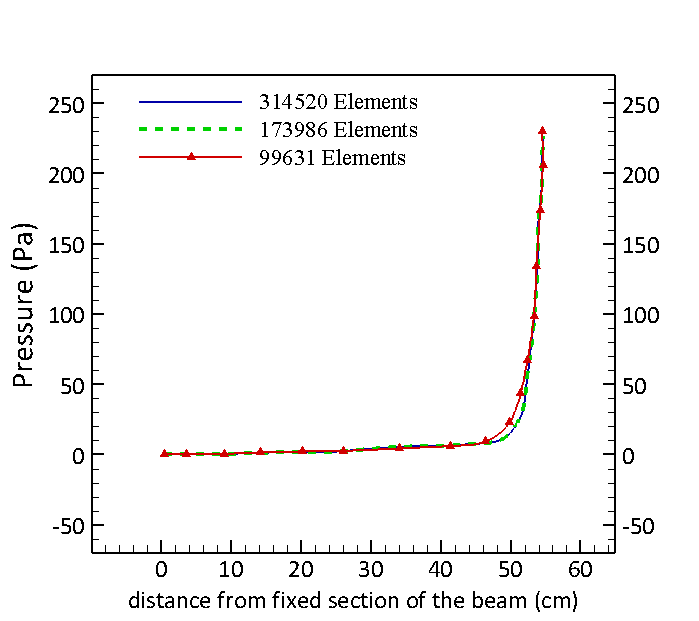}
\caption{\label{fig:pressure}The pressure distribution on the front side of the beam obtained using three computational grids.}
\end{figure}

The grid independence procedure has been carried out by use of non-uniform unstructured meshes refined near the beam and the air-water interface (Fig. \ref{fig:mesh}). In the wave-structure interaction process modeling, the whole boundary layer and the fluid-fluid interface is refined, and the sensitivity of the computation results for the pressure distribution on the beam (Fig.~\ref{fig:pressure}), the maximum deflection of the beam, and the maximum output voltage have been examined by exercising the grids with total elements of 99631, 173986, 314520 in Tab. \ref{tab.grid_maxvoltage}. These three computational grids have been generated by concentrating on refinement near the beam and the air-water interface. The results of the grid with 173986 and 314520 elements including the pressure distribution shows a favorable agreement. Also, the maximum deflection of the beam along with the maximum output voltage, the computational outcomes have been converged with less than 1\% difference. Consequently, the domain with 173986 elements was adopted as the optimum grid.

\begin{table}[htbp]
\caption{\label{tab.grid_maxvoltage} Examination of the grid independence test for the maximum output voltage}
 \begin{tabular}{c c c c}
 \hline
   Grid & Coarse & Fine & Finest \\
   \hline
   Maximum output voltage (V) & 0.343 & 0.381 & 0.385 \\
   \hline
\end{tabular}
\end{table}

It is feasible to use an adaptive time-step computation determined based on the relative and absolute tolerances. By considering the solver convergence, if larger time-steps are selected, they do not affect the final solutions because the accuracy is determined according to these tolerances. The relative and absolute tolerances are set to 0.01 and 0.001, respectively. The implicit BDF (backward differentiation formula) with manual time-stepping is applied in the fully-coupled solver, and automatic Newton settings in the solver helps prevent divergence. As seen in Tab. \ref{tab.time}, two time-steps have nearly generated the same results for two studied parameters. Thus, in order to decrease the costs of computations, the time-step size of 0.001 has been employed.

\begin{table}[htbp]
\caption{\label{tab.time}Investigation of the time-step size independence for two parameters of maximum output voltage and maximum deflection of the beam.}
 \begin{tabular}{c p{3cm} p{2.5cm} }
 \hline
   Time-step size & Maximum output voltage & Maximum deflection of the beam  \\
   \hline
   0.0005 & 0.379 & 5.59 \\
   0.001 & 0.381 & 5.65 \\
   0.005 & 0.394 & 5.02 \\
   \hline
 \end{tabular}
 \end{table}

\section{Validation/verification}
\label{very}
\begin{figure*}[htbp]
\centering
\includegraphics[width=15cm]{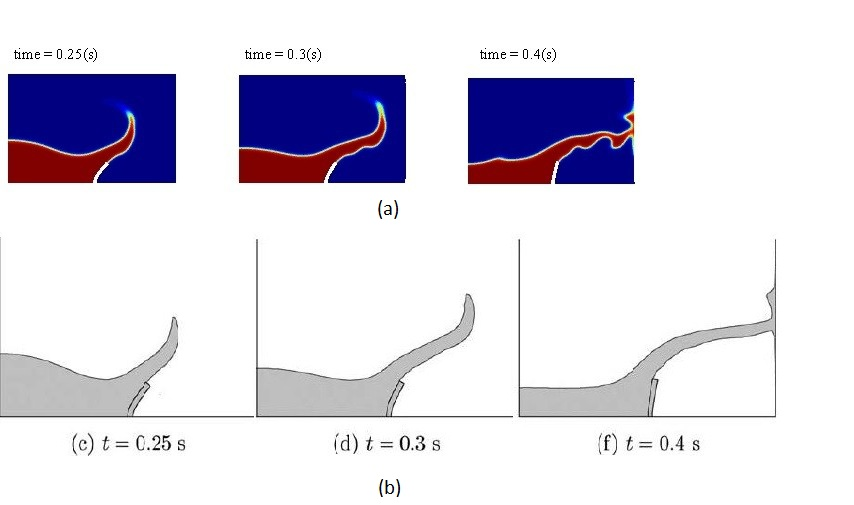}
\caption{\label{fig:volume fraction} Fluid volume contours and deflection of the beam, (a) the presented study, (b) the numerical study of \cite{walhorn2005fluid} for the dam break problem.}
\end{figure*}

To verify the accuracy of the numerical model presented in the previous section, results have been compared with Valhorne’s numerical outcomes \cite{walhorn2005fluid}. A 2D benchmark dam break involving an elastic beam was considered in Valhorne's numerical study to provide the deflection of the beam tip over time as well as the volume fraction contours. In Valhorne's study, a water column strikes an elastic beam while flowing under effect of its own weight and leads to the beam’s deflection. The interaction between the beam and fluid flow also affects the free-surface shapes.
Contours of he volume fraction of water extracted from the simulation is compared with the free-surface flow figures provided in Valhorne's numerical study. A good agreement is seen in Fig. \ref{fig:volume fraction} for the deflection of the beam and the water tongue obtained from the presented simulations and the mentioned numerical study.


\begin{figure}[t]
\centering
\includegraphics[width=9cm]{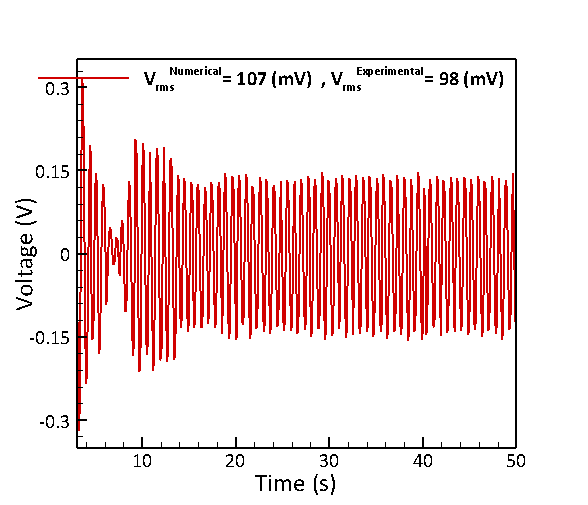}
\caption{\label{fig:voltage valid}The time-history of the voltage and comparison with the experimentally obtained RMS voltage.}
\end{figure}

Also, a laboratory set-up of the harvesting system was constructed to validate the output RMS voltage with those of the presented model. Configuration of the energy harvesting in the experiments was presented in Fig. \ref{fig:Geometry}. The piezoelectric beam with indentation of 3(cm) in water produces 76mV output RMS voltage while from the numerical simulations this value is 83 mV, which shows a 9.18\% difference as seen in Fig. \ref{fig:voltage valid}. 



\section{Results and discussion}
\label{results}
\subsection{Added mass to the tip of the beam}
The key to harvesting piezoelectric energy is the low operation frequency of this harvester owing to the frequency of surface waves at 1.37 Hz. Hence, the beam's first vibration mode must be stimulated to harness energy. Adding mass to the beam decreases the damped natural frequency, thus it is more likely to happen the resonance phenomenon. To examine the effect of adding mass to the beam tip on the output voltage, a piezoelectric beam with a thickness of 1 mm and a distance of 40 cm from the wavemaker was positioned and four masses weighing 50, 100, 130, and 180 grams were added to the beam tip. Given the constitutive relations provided in Section \ref{piezo}, the fact that the electric displacement is commensurate with the stress means that the rise in the stress range leads to generation of a higher RMS voltage. Enhancement of the displacement range of the beam tip could indicate an increase in the stress range and the generated voltage due to the state of the beam’s constraint (cantilever).

\begin{figure}[htbp]
\centering
\includegraphics[width=10cm,height=7.8cm]{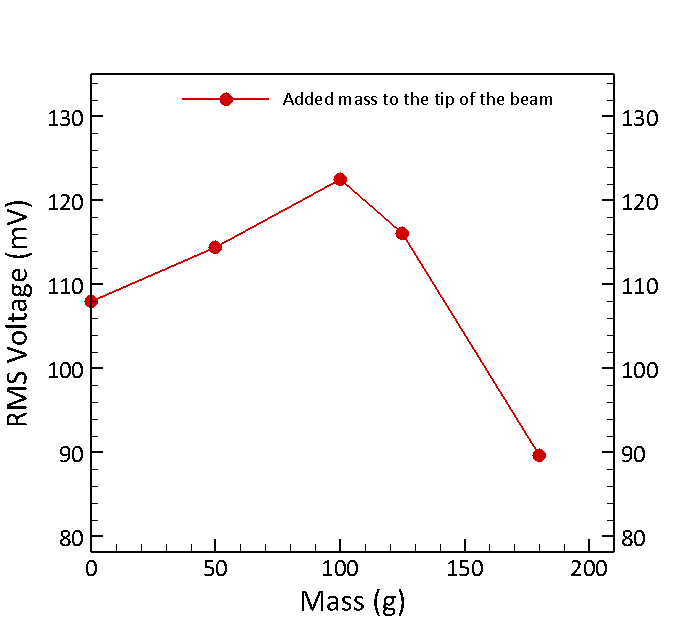}
\caption{\label{fig:mass_voltage} The RMS output voltage from the piezoelectric beam with and without the mass attached to the tip.}
\end{figure}

\begin{figure}[htbp]
\centering
\includegraphics[width=10.5cm]{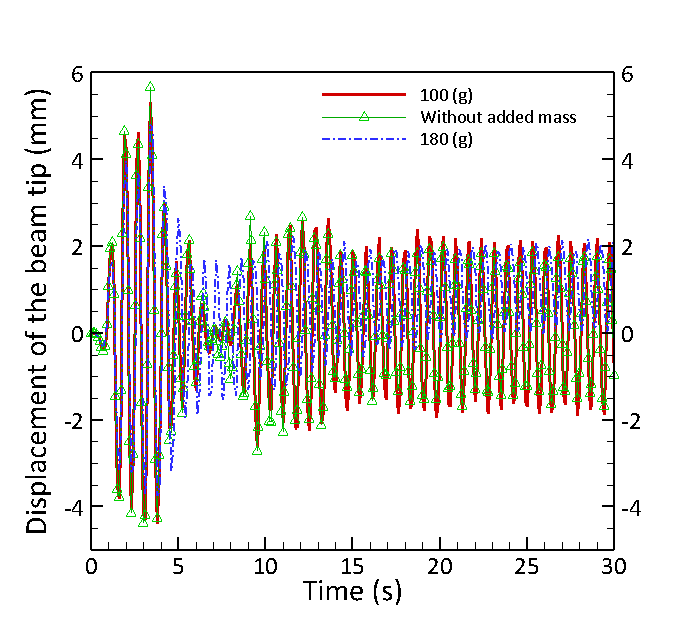}
\caption{\label{fig:mass_disp} Temporal variation of the displacement of the beam tip with and without the attached mass.}
\end{figure}

This phenomenon is important from the point of view that the tip displacement increases for the same bending force. The output voltage increases with adding the mass up to 100 grams compared to the state without the mass, and this rise is up to 13.5\% for the mass of 100 grams according to the data in Fig. \ref{fig:mass_voltage}. The rise in voltage is brought about by a decrease in the structure's damped natural frequency and its approach to the frequency of the surface waves. The voltage begins to drop for the weights between 120 grams and 180 grams as a result of moving away from the resonant frequency. The RMS voltage for the mass of 180 grams attached to the beam tip is 17\% less than the beam without the mass. As shown in Fig. \ref{fig:mass_disp}, the range of displacement for the mass of 100 grams is higher than the other states. This trend confirms the above-mentioned statements.

\subsection{The effect of the load resistance}
\begin{figure*}[htbp]
\centering
\includegraphics[width=15cm]{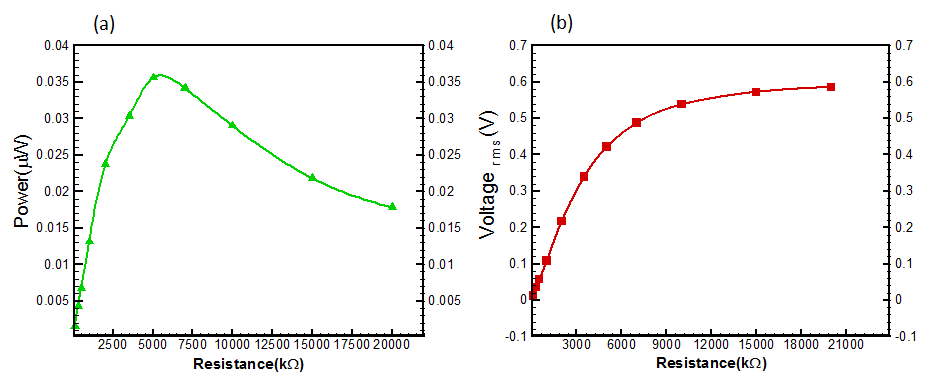}
\caption{\label{fig:Resistance_voltage} (a) The output power and (b) the RMS voltage for different load resistances.}
\end{figure*}

In all simulations, the electrical resistance of the series load with the piezoelectric converter circuit is equal to 1 $M\Omega$. By altering the load resistance value in accordance with Patel et al. \cite{patel2011geometric} work, the RMS voltage and the electric power both change. As seen in Fig. \ref{fig:Resistance_voltage}, for higher values of the load resistance, the effective voltage raises in a non-linear manner so that the slope of the graph decreases and the RMS value eventually tends to a constant value, which is justifiable due to the data obtained from Eq. (\ref{voltage}) \cite{patel2011geometric}.
\begin{equation}\label{voltage}
{U_{RL}}(t) = \frac{{(1/{R_L})}}{{1/\sqrt {{{({R_L} + {R_S})}^2} + {{(\frac{1}{{\omega C}})}^2}} }} \times {U_p}(t)
\end{equation}
where $U_{RL}$, ${R_L}$, ${R_S}$, and ${U_p}(t)$ are the generated voltage through a specified load resistance, the load resistance in series with the circuit, the internal resistance of the piezoelectric material, and the generated voltage of the harvester, respectively. However, the electric power increases up to an optimum value for higher values of the load resistance, and then begins to drop off once the optimum resistance is reached. Eq. (\ref{power}) shows that the resistance and the voltage have an inverse relationship thus, there should be an optimum resistance for the generated voltage \cite{patel2011geometric}. As seen in Fig. \ref{fig:Resistance_voltage}, the output power for 5 $M\Omega$  load resistance is 0.0356, which indicates a 2.61 times increase with respect to the reference resistance that is 1$M\Omega$ .
\begin{equation}\label{power}
 < P > {\kern 1pt} {\kern 1pt} {\kern 1pt}  = \frac{{{U^2}_{RL}(t)}}{{{R_L}}}
\end{equation}

\subsection{Inclination angle of the beam in water}
This section examines influence of the inclination angle of the beam in water on voltage generation. If the weight of the beam is substantial, the structure's orientation will play a critical role in the damped natural frequency. In the present study, the weight of the structure is not considerable, but it is possible to boost the efficacy of the transducer from the wave force by modifying the inclination (the tilt angle) of the cantilever beam in water relative to the vertical axis. In order to assess the effect of the angles on voltage, the distance from the wavemaker and the indentation of the beam in water are kept constant in comparison to the reference state.

\begin{figure*}[htbp]
\centering
\includegraphics[width=15.5cm]{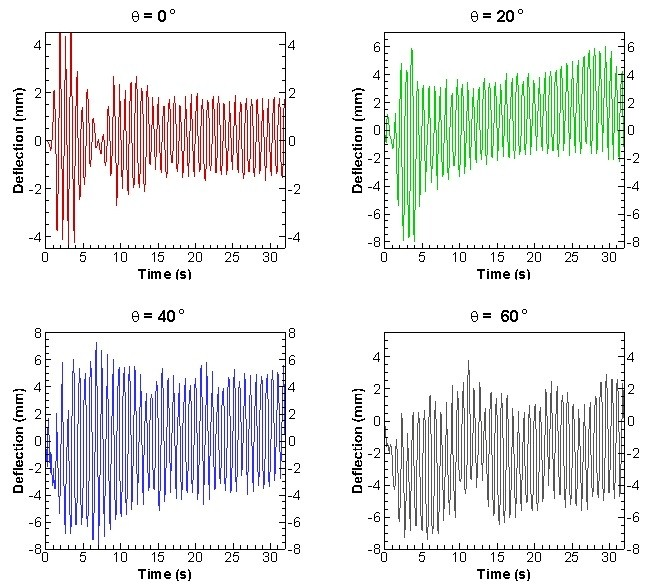}
\caption{\label{fig:deflections for abgles} Displacement of the beam tip for four different inclination angles.}
\end{figure*}

It is found that the output RMS voltage for four different inclination angles of the piezoelectric beam in water has risen relative to the vertical state (${0^{\,o}}$) so that for angles of ${20^{\,o}}$, ${40^{\,o}}$ and ${60^{\,o}}$, there are increased generated voltage of 43.2\%, 89.53\% and 68.7\%, respectively. The explanation for these increases in the RMS voltage is demonstrated in Fig. \ref{fig:deflections for abgles}, which provides the displacement of the beam tip for corresponding angles. At the angle of ${40^{\,o}}$ , the largest displacement range of any other angle is observed; furthermore , the ${60^{\,o}}$ angle possesses a displacement range  that is greater than the others and smaller than ${40^{\,o}}$. Consequently, the RMS voltage corresponding to ${60^{\,o}}$ is higher than that of two other angles but less than the voltage for ${40^{\,o}}$.

The size of structure plays a critical role in analyzing the exerted force of water waves on the piezoelectric-beam transducer. If the ratio of the structure's characteristic size (for example, its diameter) to the wave amplitude is less than 0.2, the structure is deemed small; thus, the drag and the inertial forces acting on it become significant. The forces acting on small marine structures, according to Morrison's equation, are the sum of the drag and the wave inertial forces \cite{wolfram1999alternative}, which are as follows:
\begin{equation}\label{Forces}
F = \,\,{F_{Drag}} + {F_{Inertia}} = 0.5\,\rho \,\,{C_D}\,V\,\,|V|{\kern 1pt} {\kern 1pt} \,{A_p} + {C_M}\,\rho \,\,{V_{vol}}\,\dot V
\end{equation}
\begin{equation}\label{drag and inertia}
{C_M} = {C_a} + 1
\end{equation}
where ${C_D}$, ${C_M}$, ${A_p}$, and ${V_{vol}}$ are the drag coefficient, the inertia coefficient, the projected surface area, and the volume of the structure, respectively. Both the drag and the inertial forces applied to the piezoelectric-beam transducer will change as the direction of the beam differs, but the drag force, which results from the change in the angle of inclination in the water, is not very effective in raising the voltage of the harvester. Instead, the inertial force plays a key role in increasing the forces on the beam and the range of stresses (or displacement). ${C_a}$  is the added mass coefficient, which is related to the inertia coefficient according to Eq. (\ref{drag and inertia}). This coefficient is a measure of the accelerated surrounding fluid around the beam. Fig. \ref{fig:deflections for abgles} implies that by inclining the beam, the volume of the accelerated surrounding fluid around the beam is increased relative to the reference state (${0^{\,o}}$), which leads to the growth of vibrations and the generated RMS voltage in the transducer. As a consequence, the inertia coefficient is  sensitive to the orientation of the beam.

\subsection{Thickness of the beam}
\begin{figure*}[htbp]
\centering
\includegraphics[width=15.5cm]{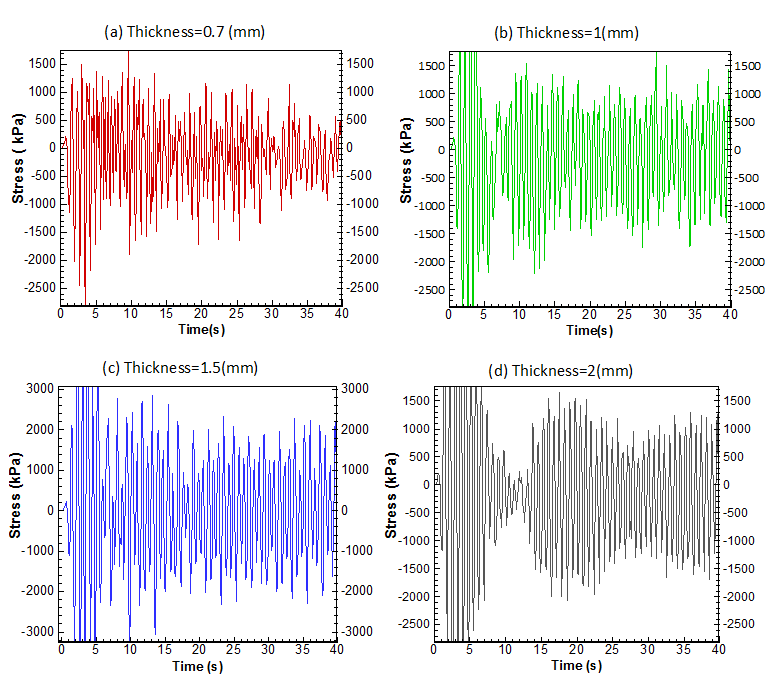}
\caption{\label{fig:stress for thickness} Temporal variations of the imposed stress in the beam corresponding to four different thicknesses.}
\end{figure*}

The impact of four different beam thicknesses on voltage generation has been studied.
By reducing the thickness of the beam from 1 mm (the reference state) to 0.7 mm, the output voltage decreases by 18.5\% while thickening by half a millimeter leads to 59.2\% rise in the output voltage. To find the reason behind this trend, it is viable to examine the stress that appears in the piezoelectric material at a point near the end of the beam. Given the obvious differences of the stress range at the end of the beam, this point is chosen to analyze the impact of thickness. Fig. \ref{fig:stress for thickness} shows that the range of stress for the beam with the thickness of 1.5 mm is greater than the other two cases, providing evidence of the origin of the rise in the output voltage. Although the state with the greatest output voltage is associated with the beam with a thickness of 2 mm, it demonstrates only a 5.81\% increase in the output voltage compared to the beam with a thickness of 1.5 mm. When the static and dynamic loadings are applied, the stress at the interface of the beam and the piezoelectric material has a direct relationship with the distance from the neutral axis of the beam, and an inverse relationship with the second moment of area of the beam cross-section. According to Fig. \ref{fig:stress for thickness}, the difference in the range of stress between these two states is not noticeable. This shows that the increase in the second moment of area of the beam cross-section has prevented the effect of increasing the distance from the neutral axis to raise the stress range and the output voltage.

\subsection{Floating harvester}
\begin{figure*}[htbp]
\centering
\includegraphics[width=13cm]{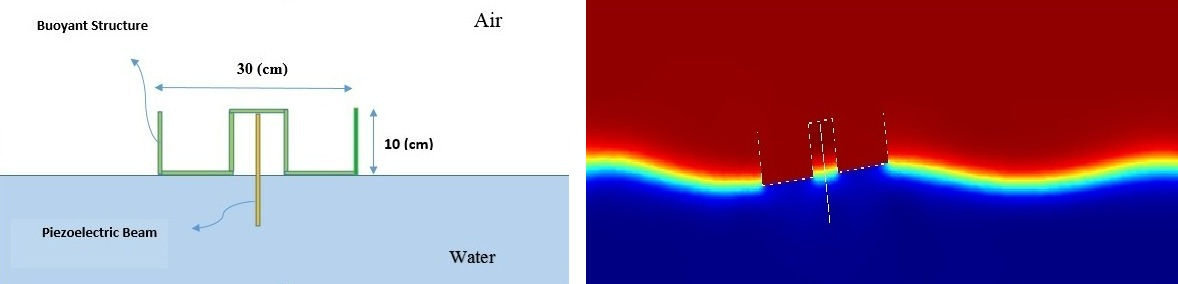}
\caption{\label{fig:schematic for indentation} The schematic view of the floating harvester (left), the free-surface interaction with the floater (right).}
\end{figure*}

\begin{figure}[htbp]
\centering
\includegraphics[width=8cm]{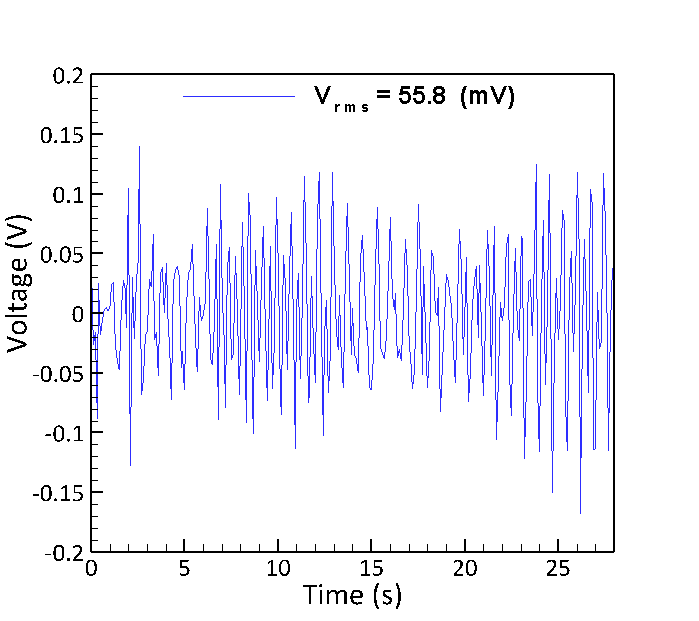}
\caption{\label{fig:time history for indentation} The time-history of the generated voltage from the floating harvester.}
\end{figure}

\begin{figure}[htbp]
\centering
\includegraphics[width=8cm]{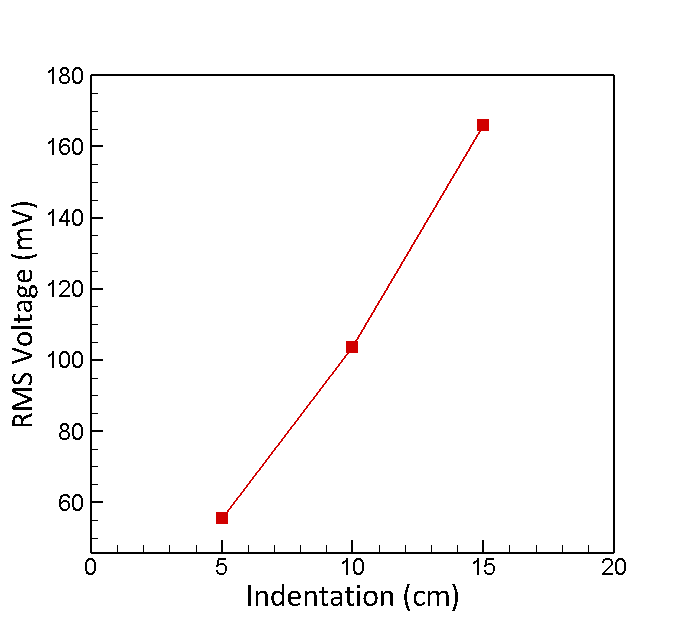}
\caption{\label{fig:voltage for indentation} The RMS of the output voltage for three indentations of the floating beam in water.}
\end{figure}

This section investigates the effect of connecting the piezoelectric beam to a floating structure with three different lengths, a schematic view as well as the free-surface interaction with the floater, and the time-history of the output voltage are given in Fig. \ref{fig:schematic for indentation} and Fig \ref{fig:time history for indentation}, respectively. According to the dimensions and the weight of the floating structure, three beams with sizes 15, 20, and 25 cm have been connected to the floating structure in the form of a cantilever beam. The RMS voltage for a 15 cm-long beam (5 cm indentation) is 55.8 mV, and as the length of the transducer beam and its indentation in the water get larger, so does the output voltage. Such that for a beam with the length of 25 cm (15 cm indentation), the output voltage is 161.94 mV, equal to an increase of 2.92 times as seen in Fig. \ref{fig:voltage for indentation}.

\begin{figure*}[htbp]
\centering
\includegraphics[width=15cm, height=7cm]{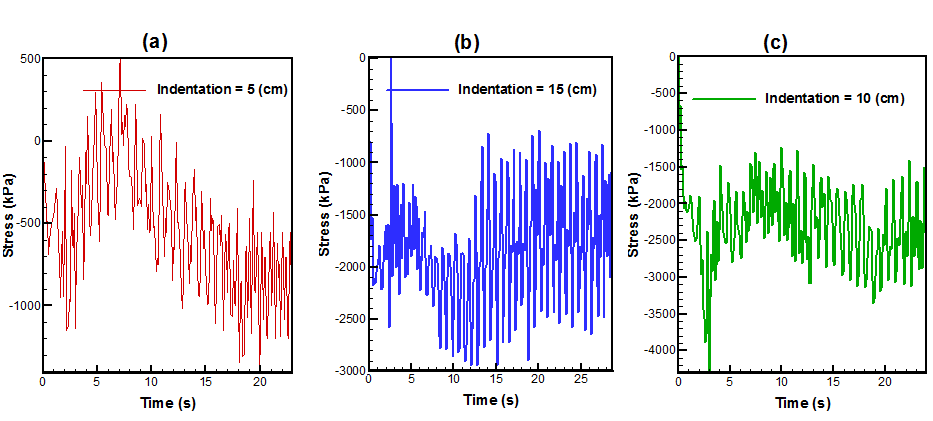}
\caption{\label{fig:stress for indentation} The imposed stress range for three indentations of the floating beam in water.}
\end{figure*}

Fig. \ref{fig:stress for indentation} reveals the origin of the trends in output power generation. It demonstrates that the lowest stress is related to the small indentations and a 15 cm indentation has a higher stress range in comparison to the other two lengths, thereby enhancement of strength of the electric field and electric displacement. It is initially assumed that the beam with smaller indentations would be stimulated more than the longer ones and a higher voltage may be generated. However, the longer the length, the less stiff the beam and the lower its first mode's natural frequency. The beam's natural frequency is inversely related to the second power of the length of the beam.

Eq. (\ref{frequency and length}) is related to the resonant frequency of a cantilever beam in which L, w, and E are the beam's length, width and the Young modulus, respectively. Also, I is the second moment of area for the beam cross-section. ${k_n}$  is equal to 3.52 for the first frequency mode.
\begin{equation}\label{frequency and length}
f = \,\frac{{{k_n}}}{{2\pi }}\sqrt {\frac{{EI}}{{w{L^4}}}}
\end{equation}

In addition, in this particular case, a longer beam means a deeper penetration of the beam inside the water, which causes the added mass to surround the beam and reduces the frequency of the first mode of the beam even more. As the first frequency mode of the beam gets closer to the excitation frequency of the water waves (which is always lower than the natural frequency of the cantilever beam) within the desired bandwidth, it is clear that the beam will get a higher output voltage with higher depth and length.

\section{Conclusion}
\label{deduction}
In this study, a  fully-coupled fluid-structure-piezoelectric model was suggested to investigate the energy harvesting from water waves. The electro-mechanical formulations and two-phase flow modeling were coupled in the framework of the finite-element approach. The prime features of the fully-coupled modeling are the ability of simulating piezoelectric harvesters at the presence of the free-surface flow (two-phase flow) and the lightweight floaters with arbitrary movements. The results are verified with the numerical study and a fabricated sample in the lab.

Therefore, following achievements can be highlighted:
\begin{itemize}
\item To approach the damped natural frequency of the structure and the excitation frequency of the water waves, a mass of 100 grams was added to the tip of the beam, and as a result, the voltage increased by 13.5\%.

\item The influence of electrical resistance on the output voltage and the power was studied. The output voltage increased when the load resistance grew until it reached a constant value. However, itbis found that the power had an optimal load resistance, for which the power enhanced by 2.61 times relative to the reference state.

\item The impact of the inertial force of water waves on the beam is accentuated by raising the inclination angle of the beam within the water relative to the vertical axis. At an angle of 40 degrees, the maximum output voltage is attained by an increase of 89.53\%.

\item The output voltage for the 1.5 mm thickness grew by 59.2 percent compared to the 1 mm thickness, while the output voltage for the 2 mm thickness increased by just 5.81 percent compared to the 1.5 mm thickness.

\item Longer indentations of the beam connected to the floater leads to the rise of the generated voltage such that the beam with the length of 25 cm (15 cm indentation) shows a 2.92 times increase in the output voltage compared to the 15 cm long beam (5 cm indentation).
\end{itemize}


\section{Acknowledgment}
This research was supported by the Iran National Science Foundation (Grant num-
ber 98017606).

\bibliographystyle{elsarticle-num}

\end{document}